\pdfoutput=1

\documentclass[runningheads]{llncs}
\usepackage{graphicx}
\usepackage{comment}
\usepackage{amsmath,amssymb} 
\usepackage{color}
\usepackage{cite}

\begin{document}
\pagestyle{headings}
\mainmatter
\def\ECCVSubNumber{5539}  

\title{Rotaflip: A New CNN Layer for Regularization and Rotational Invariance in Medical Images} 

\titlerunning{ECCV-20 submission ID \ECCVSubNumber} 
\authorrunning{ECCV-20 submission ID \ECCVSubNumber} 
\author{Anonymous ECCV submission}
\institute{Paper ID \ECCVSubNumber}

\titlerunning{Rotaflip: A New CNN Layer for Regularization and Rotational Invariance}
%
\author{Juan P. Vigueras-Guillén\inst{1} \and 
	Joan Lasenby\inst{2} \and 
	Frank Seeliger\inst{1}} 
\authorrunning{Vigueras-Guillén et al.}
%
\institute{CVRM Safety, Clinical Pharmacology and Safety Science, R\&D, AstraZeneca, Gothenburg, Sweden, \and
	Department of Engineering, University of Cambridge, Cambridge, UK.\\
	\email{JuanPedro.ViguerasGuillen@astrazeneca.com}}

\maketitle

\begin{abstract}
Regularization in convolutional neural networks (CNNs) is usually addressed with dropout layers. However, dropout is sometimes detrimental in the convolutional part of a CNN as it simply sets to zero a percentage of pixels in the feature maps, adding unrepresentative examples during training. Here, we propose a CNN layer that performs regularization by applying random rotations of reflections to a small percentage of feature maps after every convolutional layer. We prove how this concept is beneficial for images with orientational symmetries, such as in medical images, as it provides a certain degree of rotational invariance. We tested this method in two datasets, a patch-based set of histopathology images (PatchCamelyon) to perform classification using a generic DenseNet, and a set of specular microscopy images of the corneal endothelium to perform segmentation using a tailored U-net, improving the performance in both cases. \footnote{This work was originally submitted to ECCV 2020.}

\keywords{histopathology, microscopy images, denseNet, U-net}
\end{abstract}

\section{Introduction}

In recent years, deep learning (DL) have gained large popularity in the field of medical image analysis \cite{Litjens17}. DL, specifically convolutional neural networks (CNNs), allows to automate diagnostic tasks in images, such as tissue segmentation, image classification, or cell counting. CNNs are mainly comprised of convolutional layers, which are filters that execute operations in images in a convolutional manner. One property of convolutions is their translation equivariance, that is, patterns that appear at different spatial positions are encoded in the same feature maps extracted by the convolutional layers. Therefore, CNNs are highly efficient at sharing parameters, significantly reduce overfitting, and can be used for very large images without increasing the number of parameters. However, CNNs are not invariant to other common transformations, such as rotation and reflection. This is usually tackled with data augmentation, performing those transformations to the input image and, thus, increasing the number of training examples. Although this technique can achieve certain level of rotation invariance, each rotated image is treated independently and they can lead to different CNN outputs. 

These limitations are particularly relevant in many medical images where there is not a clear orientation, i.e. there does not exist an up, down, left, or right. Since CNNs do not exploit these rotational symmetries, the performance is usually suboptimal. For instance, Veeling \textit{et al.} \cite{Veeling18} indicated that CNNs exhibit predictions with erratic fluctuations for histopathology images under input rotation and reflection.

Recent publications have proposed different approaches to enforce rotational invariance in CNNs. The common idea in these methods is to apply rotations and reflections within the network, either to the feature maps or the convolutional filters. However, rotations would require to perform interpolation, which introduces many problems, such as computational complexity, distortions, and the need for zero-padding, unless they were done over angles of $k \cdot 90^{\circ}, k \in \{0,1,2,3\}$. In this respect, all the following publications only used those four types of rotations when referring to this transformation (and so we will do in this paper). For instance, Dieleman \textit{et al.} \cite{Dieleman15} presented a CNN architecture for classification of astronomy images where the input image was rotated, flipped, and cropped to yield different viewpoints, subsequently each viewpoint was processed by the same stack of convolutional layers, and finally their output representations were concatenated and processed by a dense layer to obtain a prediction. This work was further extended to propose new CNN layers to generate, stack, and/or combine rotated copies of a set of inputs \cite{Dieleman16}. In contrast, Gao and Ji \cite{Gao17} suggested to perform the rotational and flipping transformations to the convolutional filters, creating four (only rotations) or eight (rotations and reflections) filters per original filter in a convolutional layer, yielding a different feature map for each new filter, and performing max-out among the four or eight transformed maps to reduce it to one single map. Cohen and Welling \cite{Cohen16} proposed a similar idea, but instead of creating several new filters per original convolutional filter, the convolutions would be performed in groups, that is, a single filter would be transformed to perform four or eight different convolutions and the results would be combined to generate a single feature map. This new type of convolutional layers, named G-CNN, has been quite successful: Weiler \textit{et al.} \cite{Weiler18} used it to build steerable filters, where the filter weights were shared to improve generalization; Veeling \textit{et al.} \cite{Veeling18} applied it for the classification of pathology images; and Graham \textit{et al.} \cite{Graham19} used it in their network Rota-Net to segment colon histology images.

In contrast to CNNs, neural networks (NNs) are prone to overfitting and, thus, regularization methods are usually necessary. In the context of medical images, NNs (also called fully connected or dense layers) are added at the end of CNNs to perform classification or regression. In those cases, the most common regularization method is the inclusion of dropout layers in between dense layers \cite{Srivastava14}. Briefly summarized, a dropout layer will remove a certain number of neurons (randomly selected) at each training iteration. This forces the remaining units to avoid to rely on a few specific previous neurons (co-adaptation) as they might be occasionally excluded, hence distributing its knowledge among more neurons and making the model more robust. Although uncommon, dropout layers can also be used in between convolutional layers, which set to zero a certain percentage of pixels in the feature maps. However, dropout might become ineffective since the CNNs' gradients are averaged over the spatial extent of the feature maps and, thus, different dropout patterns might not have an effect on the averaged gradients or might distort in excess if the drop rate is too high. Furthermore, batch normalization (BN) layers, commonly used before convolutional layers, already provide certain level of regularization \cite{Ioffe15}. Nevertheless, there might be cases where further regularization can be beneficial, although this is a complex scenario where multiple variables exist: the complexity and amount of data, the task to perform, the design of the network (depth, width,...), etc. 

In this work, we propose a new CNN layer, named \textit{rotaflip}, that performs regularization while providing a certain degree of rotation invariance. This new layer can be seen as an alternative to dropout layers in the convolutional part of a CNN, and provides better performance in comparison to aiming for rotation invariance through data augmentation. We experimented with two public datasets: (1) a patch-level set of histopathology images, named PatchCamelyon \cite{Veeling18}, to perform classification using a DenseNet-121 \cite{Huang17}; and (2) a set of \textit{in vivo} specular microscopy images of the corneal endothelium \cite{ViguerasBMC19} to perform cell segmentation using a U-net \cite{Ronneberger15} and a DenseUnet \cite{ViguerasSPIE19}. It is important to highlight that it is not our goal to surpass the state-of-the-art in any of the datasets here described but to prove that the inclusion of rotaflip layers in generic networks could improve the performance.

The contributions of this work are as follows: (1) we propose a novel regularization method that exploits the rotational symmetries in medical images; (2) we evaluate the method in two different datasets with two generic networks, hence showing the method is not problem, data, or network dependent; (3) we compare our method with dropout, proving that it performs similarly or better, depending on the problem to solve.


\section{Methods}

\subsection{Rotaflip Layer: Definition and Motivation}

A rotaflip layer can be interpreted as a way of regularizing a CNN by performing random rotations and reflections to the feature maps, which in return also provides a certain level of rotation invariance. As mentioned above, the idea of performing such transformations within the network has been previously proposed, but in those cases all possible changes were applied to each feature map and combined with the original maps in order to avoid the exponential growth of channels. Furthermore, those methods aimed for rotational invariance and regularization was never studied. In contrast, our work changes that idea to simply perform a single transformation to a small percentage of maps, without keeping the original orientation of the converted maps, and being the transformation of each map independent. Specifically, a single transformation is randomly selected among the eight possibilities (four rotations and two reflections) for each map to convert. Therefore, our layer does not increase the size of the feature maps; it just shuffles the orientation of a few ones. Like dropout, the transformations are only performed during training, and there is only one hyperparameter to tune: the percentage of maps to transform.

We believe that rotaflip layers are more powerful than dropout layers when applied in the convolutional part of CNNs because dropout introduces noise patterns that do not provide any actual knowledge (beyond artificially increasing the training data with unrepresentative examples), whereas rotaflips add the rotated/reflected versions of real patterns which induces the network to learn them.

To the best of our knowledge, this idea has not been tested yet, probably due to its restricted use to images without orientation. Indeed, natural images usually possess an orientation that is sometimes critical for its identification (letters, numbers, etc.), and therefore these transformations could be counterproductive.

\subsection{Materials and Network Architectures}

\subsubsection{PatchCamelyon \& DenseNet.}
\begin{figure}[t]
	\centering
	\includegraphics[width=1\linewidth]{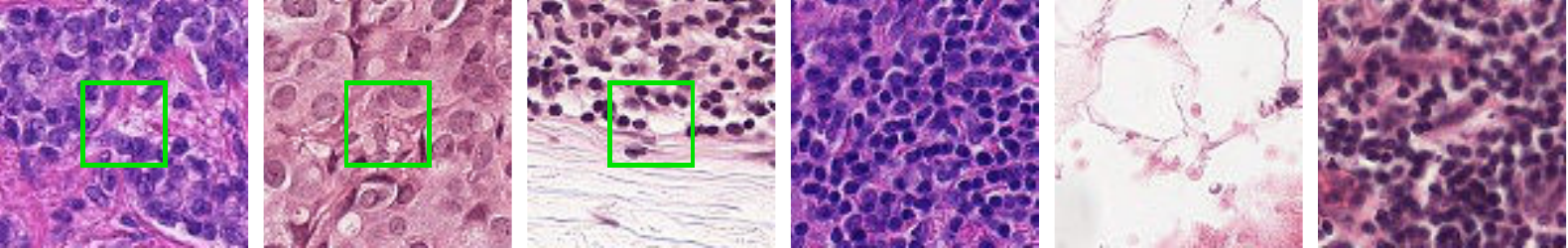}
	\caption{Six representative examples of the PatchCamelyon dataset; from left to right, three positives (with a green square of 32$\times$32 pixels), and three negatives patches. The green square indicates the area used to determine the class, this is, a patch belongs to the positive class only if any pixel within the square was annotated as cancer pixel.}
	\label{fig01}
\end{figure}

The main dataset in this work is the PatchCamelyon \cite{Veeling18}. This is a set of 327,680 color patches (96$\times$96 pixels) extracted from 400 H\&E stained whole-slide images of sentinel lymph node sections \cite{Bejnordi17}, where each patch is annotated with a binary label indicating the presence of metastatic tissue. Specifically, a positive label indicates that the center 32$\times$32 pixels in the patch contains at least one cancer pixel (Fig.~\ref{fig01}). The authors already divided the set into 10 folds, setting eight folds for training (262,144 examples), one for validation (32,768 examples), and one for testing. According to the authors, there is no overlap between patches and all folds have a 50/50 balance between positive and negative examples. 

To solve the problem, the PatchCamelyon authors first adapted and reduced a DenseNet \cite{Huang17} such that there were 5 dense blocks with only one convolutional block in each. This network had 128K parameters and yielded an accuracy of 87.20\%. Subsequently, they substituted the convolutional layers for G-CNNs \cite{Cohen16}, obtaining a boost in accuracy up to 89.80\% ($\text{AUC}=96.3$) \cite{Veeling18}. Similarly, Kassani \textit{et al.} \cite{Kassani19} reached an accuracy of 87.84\% with the predefined DenseNet-201 \cite{Huang17}, which has 4 dense blocks containing 6, 12, 48, and 32 convolutional blocks, respectively, with a total of 18.3M parameters. This clearly shows the main advantage of DenseNets, which is its strength against overfitting when it is enlarged with many more convolutional layers to yield a slightly higher accuracy. It is worth noting that a DenseNet does not have any hidden NN layer at the end of the network; instead, it computes the average value per feature map in the final set of maps and it fully connects them to the output nodes (2 nodes in this problem).

\begin{figure}[t]
	\centering
	\includegraphics[width=0.65\linewidth]{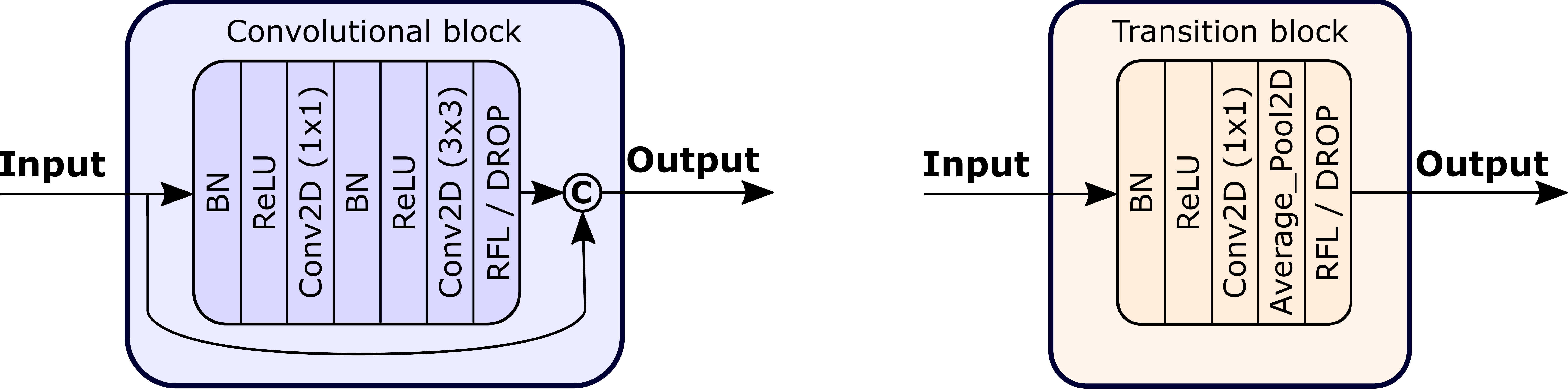}
	\caption{(Left) One convolution block in the DenseNet. The first convolutional layer, Conv2D(1$\times$1), is used for feature reduction (to $4k$ maps) and the second yields $k$ feature maps, which are then concatenated to the input features. (Right) One transition block. In both cases, the rotaflip layer (RFL) or dropout layer (DROP) is placed at the end.}
	\label{fig02}
\end{figure}

In this work, we were limited by a GPU with 12GB of memory, which cannot fit a DenseNet-201 with rotaflip layers. Therefore, we used the smaller DenseNet-121 instead \cite{Huang17}, which has 4 dense blocks with 6, 12, 24, and 16 convolutional blocks, respectively, and a total of 7M parameters. In all cases, the DenseNets had a growth rate of $k=32$ (the number of output feature maps in each convolutional layer). We used the code implementation created by the Keras team \cite{KerasTeam}. In our experiments, we added a rotaflip layer (or a dropout layer) at the end of each convolutional and transition block (Fig.~\ref{fig02}). The remaining hyperparameters were: batch size of 128 with equal number of positive and negative patches, binary cross-entropy as loss function, nadam optimizer \cite{Dozat2016}, 300 epochs, a initial learning rate of $lr=0.001$, and a rate decay of $lr_{decay}=0.99$, being the updated rate at each new epoch $lr_{new}= lr\times (lr_{decay})^{epoch}$.

\subsubsection{Endothelium \& U-net/DenseUnet.}
\begin{figure}[t]
	\centering
	\includegraphics[width=0.65\linewidth]{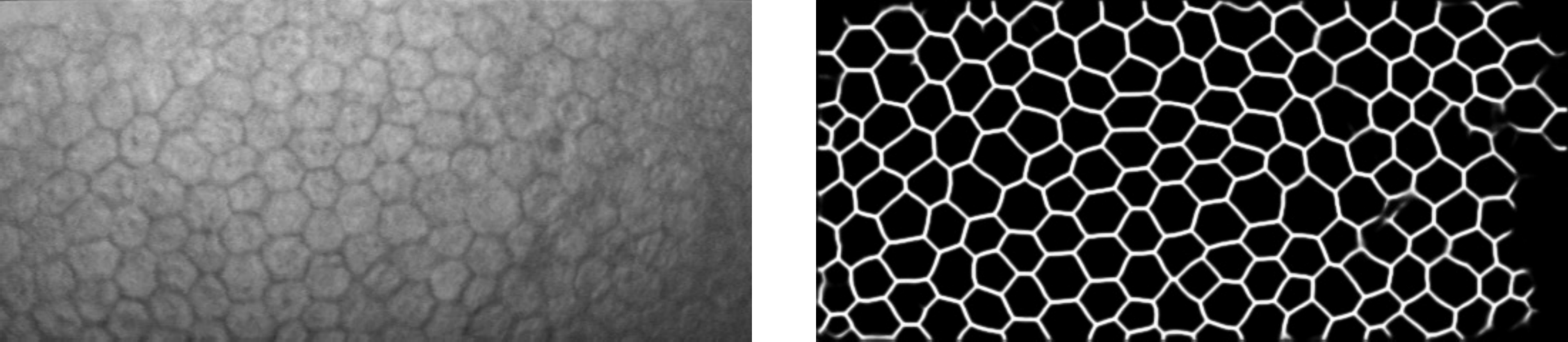}
	\caption{(Left) One representative example of the endothelium dataset. (Right) Cell segmentation yield by the U-net.}
	\label{fig03}
\end{figure}

To evaluate our method in Fully Convolutional Networks (FCNs), we used a public dataset of corneal endothelium images from the Rotterdam Eye Hospital \cite{ViguerasBMC19}. This set is comprised of 50 gray-scale specular microscopy images (240$\times$528 pixels) of endothelial cells (Fig.~\ref{fig03}). The dataset's authors designed a U-net to detect the cell edges, using the accuracy at the pixel level as metric, and they discussed the need for regularization. Indeed, their U-net was comprised of four resolution blocks (in both, the down- and up-sampling path) where each block contained two convolutional layers with a dropout layer of a 50\% drop rate in between \cite{ViguerasBMC19}. 

In this work, we first created two images of 240$\times$240 pixels per each original image so that the eight possible transformations of the rotaflips could be used. The resulting set of 100 images was subdivided in five folds, using four for training and one for testing. We also reduced the depth of the deepest layers to accommodate the limited resources, setting the number of filters to 32, 64, 96, and 128 for the four resolutions blocks, respectively, and a total of 1.8M parameters (originally, the last two blocks had 128 and 256 filters). The original paper used the two types of reflections as data augmentation; here, we included the reflections and the four rotations as data augmentation. In our experiment, we tested the original U-net without the dropout layers and three alternatives: one with the dropout layers as designed by the dataset's authors, another with the rotaflip layer at the end of each resolution block and no dropout, and a last one where both, dropout and rotaflip, were employed together. We used the same hyperparameters aforementioned in the DenseNet, with the exception of a batch size of 4 images. 

In a later work, the dataset's authors increased the set up to 140 images and updated their network to a DenseUnet \cite{ViguerasSPIE19}, reducing the dropout to 20\%, and in a subsequent work they further enlarged the set up to 738 images \cite{ViguerasEMBC19} where dropout was no longer necessary since no overfitting occurred. To evaluate whether rotaflip layers are detrimental in networks where overfitting is nonexistent, we also replicated their latest DenseUnet network (6.9M parameters), following the same approach as detailed above, and we experimented with their largest dataset, which we converted into 1476 images of 240$\times$240 pixels.

\section{Experiments}

\subsection{PatchCamelyon}

\begin{figure}[t]
	\centering
	\includegraphics[width=1\linewidth]{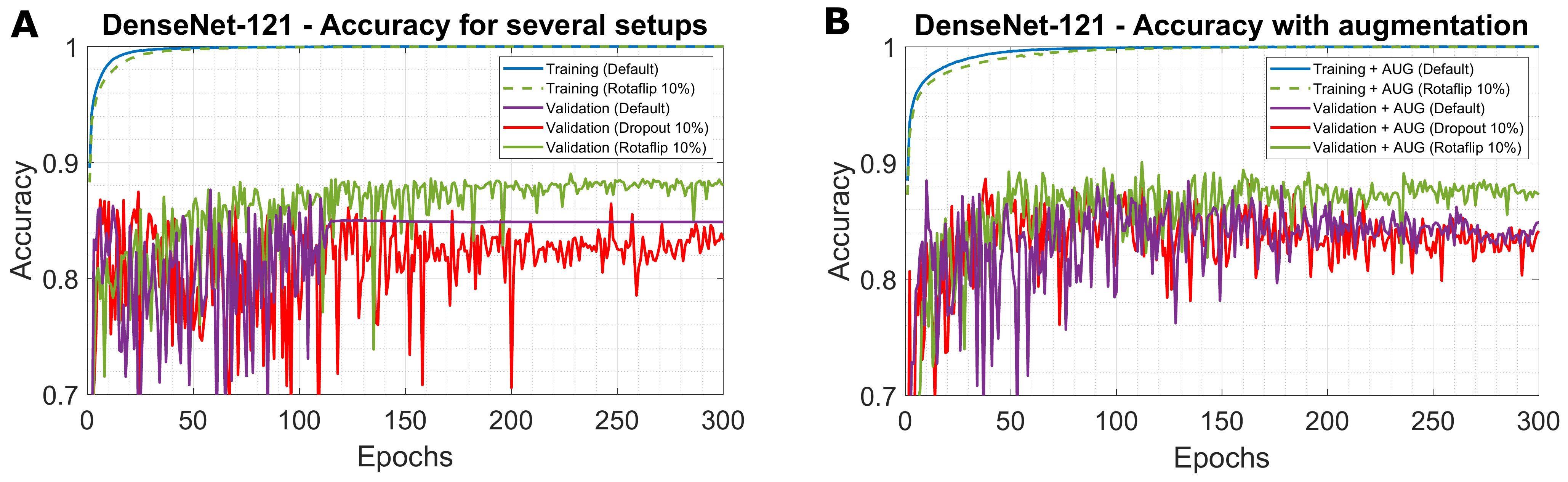}
	\caption{DenseNet-121. Accuracy in the training and validation sets for the default DenseNet-121 and for the DenseNet-121 with a dropout layer of 10\% rate or a rotaflip layer with 10\% rate at the end of each convolutional and transition block (as in Fig.~\ref{fig02}). \textbf{(A)} Without data augmentation in the input. \textbf{(B)} With data augmentation, which consist of the eight transformations (rotations and reflections). }
	\label{fig04}
\end{figure}

First, we evaluated the default DenseNet-121 and we observed a perfect fit in the training set (100\% accuracy, Fig.~\ref{fig04}-A) and a convergence in the validation set at around 120 epochs with 84.89\% accuracy, but no accuracy drop was observed in the validation set and the convergence value was in accordance with the results provided by Veeling \textit{et al.} \cite{Veeling18} and Kassani \textit{et al.} \cite{Kassani19}, where the former had a much smaller DenseNet and the latter a deeper one. Thus, the network was clearly overlearning the details of the training data but still generalized relatively well and as expected. The experiment was repeated to ensure the behavior was not random, and we observed the same convergence (at epoch 135 this time) with an accuracy 84.73\%.

Subsequently, we added a rotaflip layer at the end of each convolutional and transition block (as depicted in Fig.~\ref{fig02}), and we repeated the experiment with different rotaflip rates. Similarly, the same experiments were done with dropout layers instead of rotaflip layers. In all cases, these two regularization methods prevented the network to converge to a specific point because of their randomness nature, but they all tended towards certain range of values and still no sudden accuracy drop was observed in any case (the example with 10\% rate is depicted in Fig.~\ref{fig04}-A). More importantly, the dropout setup was degrading the performance with just a small drop rate, whereas the rotaflip setup increased the accuracy and reduced the fluctuations (Fig.~\ref{fig04}-A). In all cases, the network kept overfitting the training set to a 100\% accuracy, although more epochs were needed for this event to happen. 

We then repeated the same experiment but adding data augmentation in the input, being the augmentation simply the eight described transformations (no other type of augmentation --such as noise addition, deformations, gamma distortion, etc.-- were included in order to isolate the rotational/reflecting transformations). The main differences were the lack of total convergence in the default network and the reduction of fluctuations in the dropout experiment with lower rate (Fig.~\ref{fig04}-B) --not so in the higher rates of dropout--. Indeed, data augmentation was slightly improving the performance in certain cases (Table~\ref{table_DenseNet}), but its main benefit was providing rotational invariance (analyzed below). 

\begin{figure}[t]
	\centering
	\includegraphics[width=0.60\linewidth]{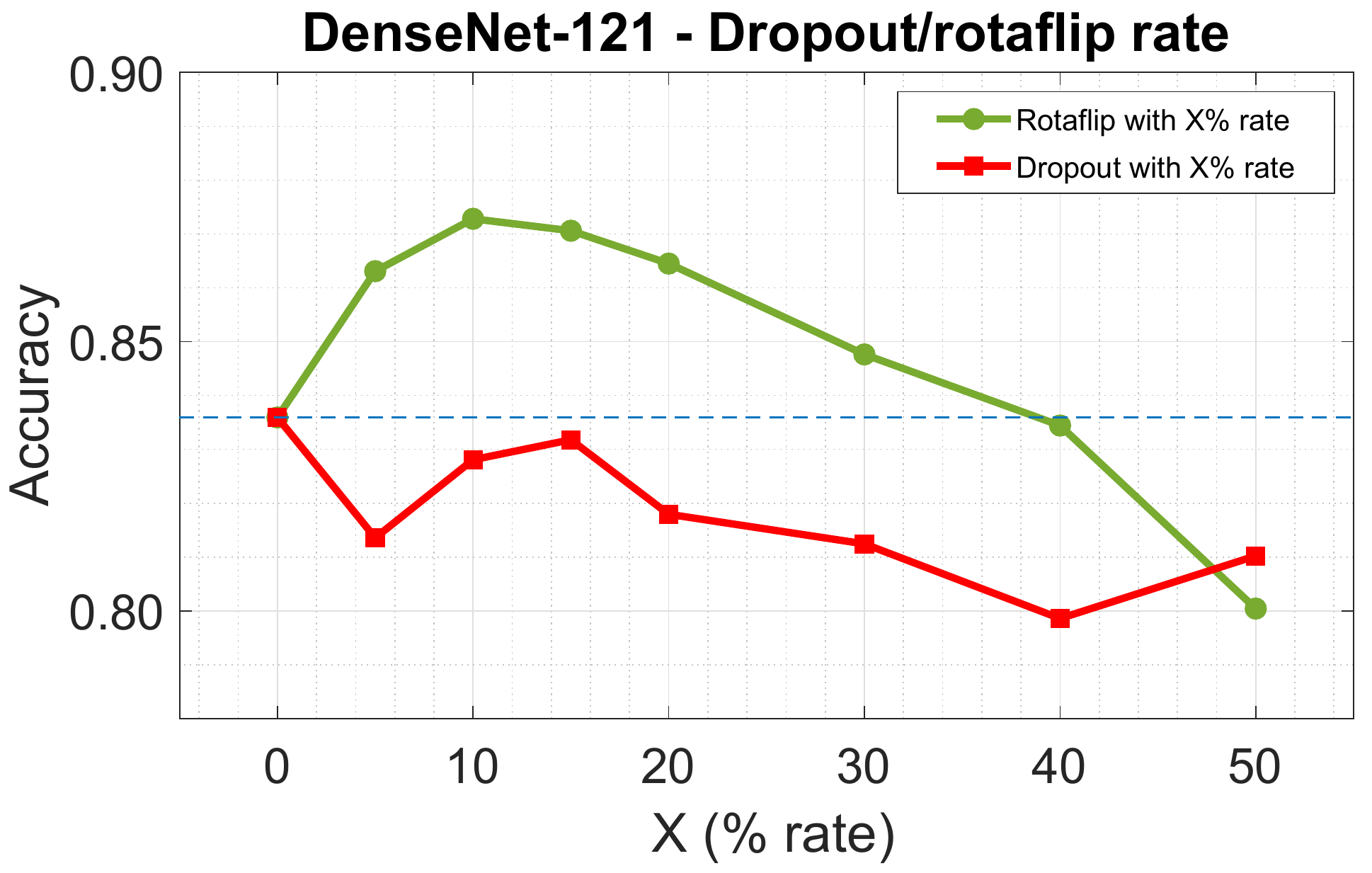}
	\caption{DenseNet-121. Accuracy in the validation set (including the eight transformed versions for each image) for different rates of dropout and rotaflip (0\% rate indicates the default DenseNet-121) and no data augmentation. Accuracy is computed as the average among the last 10 epochs (epochs \#291--300).}
	\label{fig05}
\end{figure}

To study the relevance of the rotaflip and dropout rate and to avoid any bias due to accuracy fluctuations, we computed the average accuracy in the validation set for the last 10 epochs in each setup. For this, we included the eight transformed versions of each image in the validation set (the graphs in Fig.~\ref{fig04} were obtained with only the default orientation of the validation images). This experiment showed how the accuracy decreased in all the dropout cases, whereas a boost in performance was observed for small rates of rotaflips (Fig.~\ref{fig05}).

The previous analysis considered the last epochs to compute the metrics in an attempt to make a fair comparison. However, it is a common practice to keep the model with the highest validation accuracy for further analysis in the test set since there might be some degree of overfitting after certain number of epochs. Given the large fluctuations in accuracy (Fig.~\ref{fig04}), it seemed counterproductive to perform exactly so, as this might not give a correct overview of the generalization properties of the model (i.e. a good fit in the validation set might be a coincidence and thus not extrapolable to the test set, and vice versa, particularly in early epochs). To minimize this bias, we computed the accuracy for the eight transformed versions of each image in both, the validation and test sets, for the best model according to the validation set, and we reported the average value (Table~\ref{table_DenseNet}). We observed that the accuracy obtained in the test set was systematically lower than in the validation set in all cases, but since we did not tune the hyperparameters of the network, it was clear that we were not overfitting to the validation images; the test set was simply more complex. Nevertheless, the same pattern was observed, being the rotaflip with 10\% rate the best model (Table~\ref{table_DenseNet}). Furthermore, we believe that the peak observed in the case of dropout 40\% was a simple coincidence due to the aforementioned issue with the fluctuations (higher dropouts generated larger fluctuations, thus corroborating that dropout was not appropriate for this dataset and network).

\setlength{\tabcolsep}{4.5pt}
\begin{table}
	\begin{center}
		\caption{DenseNet-121. Accuracy in the test+validation set (including the eight transformed versions for each image) for the best model in each setup. Each column indicates the drop/rotaflip rate, being the default DenseNet-121 the one with rate 0\%. In bold, the best model.}
		\label{table_DenseNet}
		\begin{tabular}{lrrrrrrrr}
			\hline\noalign{\smallskip}
			& 0\% & 5\% & 10\%  & 15\% & 20\% & 30\% & 40\% & 50\%  \\
			\noalign{\smallskip}
			\hline
			\noalign{\smallskip}
			Rotaflip  & 82.76 & \textbf{86.10} & 85.97 & 85.00 & 84.94 & 82.64 & 82.03 & 80.87  \\
			Dropout   & 82.76 & 83.14 & 82.85 & 84.75 & 82.09 & 81.93 & 84.33 & 81.02  \\
			Rotaflip + AUG  & 84.43 & 84.14 & \textbf{85.42} & 85.31 & 85.10 & 82.48 & 82.03 & 82.58  \\
			Dropout + AUG   & 84.43 & 82.98 & 83.86 & 79.90 & 83.50 & 80.90 & 80.30 & 80.76  \\
			\hline
		\end{tabular}
	\end{center}
\end{table}

Finally, to evaluate the robustness against rotations and reflections, we measured the agreement in classification (in percentage) for the different versions of an image, regardless the true label of the image, and we called this metric \textit{agreement}. For example, given the eight versions of an image, a model that classifies two versions as positive and six as negative has an \textit{agreement} $= (6/8)\cdot 100\% =75\%$ for that image, being the model \textit{agreement} the average among all images. Thus, a model with high \textit{agreement} could be interpreted as rotationally robust. 

Similar as previous experiments, we averaged the metric among the last 10 epochs. The default network had an agreement of 93.78\%, which increased to 97.81\% with data augmentation. Similarly, by adding data augmentation, the Dropout-10\% model changed from an agreement of 94.27\% to 97.85\%, and Rotaflip-10\% increased from 93.81\% to 98.04\%. As indicated in Table~\ref{table_DenseNet} and observed in Fig.~\ref{fig04}, data augmentation along with rotaflips did not improve the accuracy but it helped the network to be more consistent with the classification of the different versions of the images.

\subsection{Corneal Endothelium}

\begin{figure}[t]
	\centering
	\includegraphics[width=0.60\linewidth]{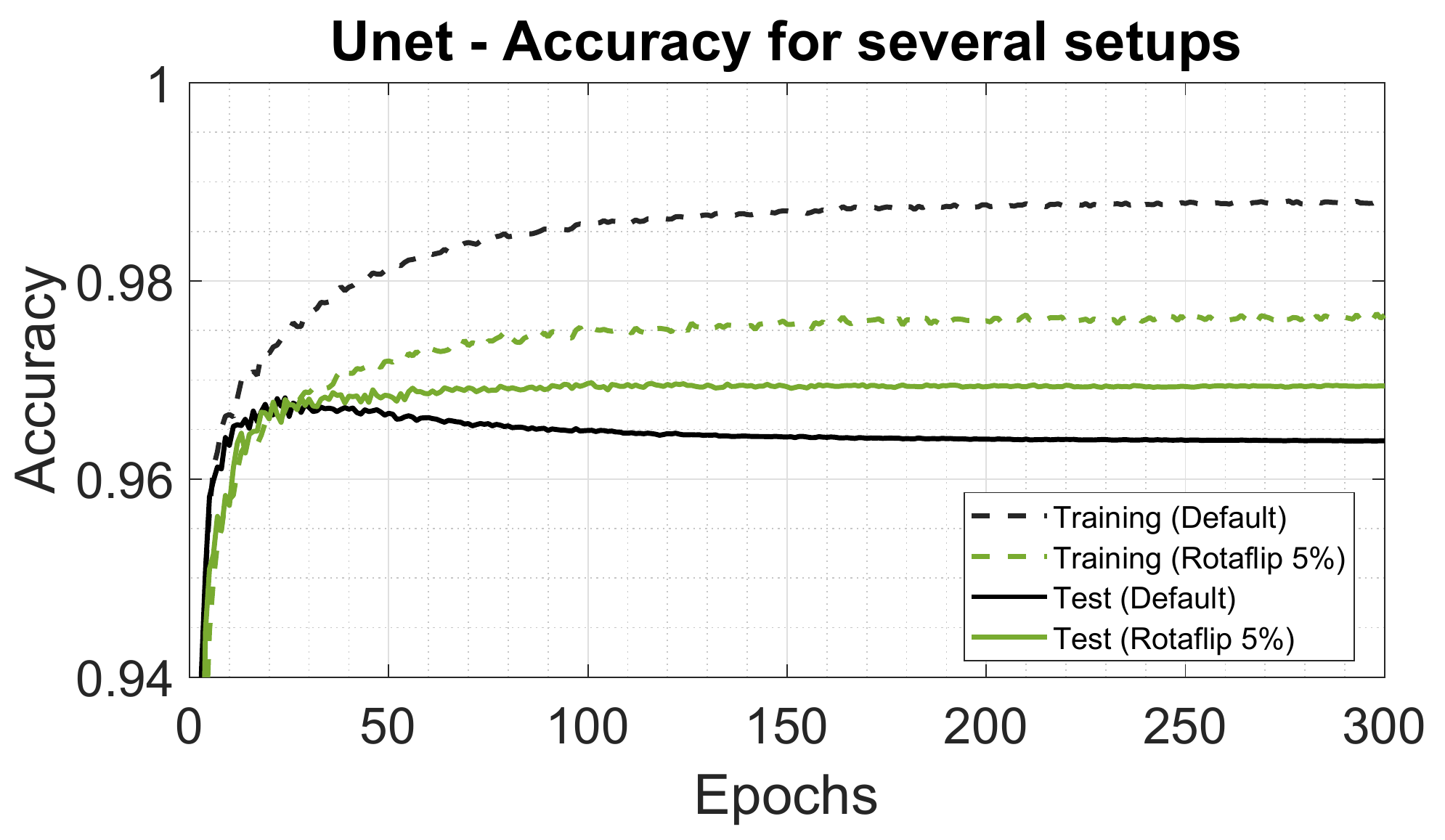}
	\caption{Unet (set of 100 endothelial images). Accuracy in the training and test sets for the default U-net (black) and the U-net with rotaflip layers of 5\% rate (green).}
	\label{fig06}
\end{figure}

The default U-net (without dropout layers) with the set of 100 images displayed a small overfitting (Fig.~\ref{fig06}), similar to the experiments presented by the dataset's authors \cite{ViguerasBMC19}. Both regularizers, dropout and rotaflip, reduced the overfitting and improved the performance (Table~\ref{table_Unet}). Overall, we observed that only very small rotaflip rates were optimal, whereas dropout was less sensitive to small rate changes and yielded a higher accuracy. Qualitatively, we did not observe meaningful differences in the segmentation yielded by the different setups. Interestingly, if both regularizers were used together, a peak in accuracy was observed in approximately their respective best rates (Table~\ref{table_Unet}), suggesting that these regularizers act independently and are not interrelated. Finally, it is worth noting that our replicated U-net did not require a 50\% dropout to achieve the best performance, as the original authors indicated, probably due to the reduction of feature maps in the deepest resolution blocks.

In the alternative experiment, the DenseUnet \cite{ViguerasSPIE19} with the set of 1476 images showed almost no indication of overfitting, where both lines, training and test, converged together to close values (training accuracy of 96.89\% and a test accuracy of 96.72\%). Nevertheless, the best setup with rotaflip layers (2.5\% rate) increased the test accuracy up to 96.74\%, whereas dropout could not improve the test accuracy in any case. Given the very small increase in accuracy by rotaflips, it could not be concluded with certainty that rotaflips were beneficial in the absence of overfitting, although they were certainly not detrimental for the network.

\setlength{\tabcolsep}{4pt}
\begin{table}
	\begin{center}
		\caption{Unet (set of 100 endothelial images). Accuracy in the test set for different setups, including the default Unet (rate of 0\%) and the inclusion of rotaflip, dropout, or both (Rot+Drop), with the indicated rates. In bold, the best accuracy for each setup.}
		\label{table_Unet}
		\begin{tabular}{lrrrrrrrrr}
			\hline\noalign{\smallskip}
			 & 0\% & 2.5\% & 5\% & 7.5\% & 10\% & 12.5\% & 15\% & 17.5\% & 20\% \\
			\noalign{\smallskip}
			\hline
			\noalign{\smallskip}
			Rotaflip  & 96.39 & 96.79 & \textbf{96.94} & 96.82 & 96.82 & 96.80 & 96.69 & 96.60 & 95.47 \\
			Dropout   & 96.39 & 96.76 & 96.89 & 96.87 & 96.93 & 96.97 & \textbf{97.05} & 97.01 & 96.97 \\
			Rot+Drop  & 96.39 & 96.81 & 96.89 & \textbf{96.93} & 96.90 & 96.80 & 96.91 & 96.74 & 96.48 \\
			\hline
		\end{tabular}
	\end{center}
\end{table}

\section{Discussion}

In this work, we have presented a method to add some regularization in deep CNNs by simply rotating and/or reflecting a few randomly selected feature maps after each convolutional layer. We have shown that this method boosts the performance on medical images where orientational symmetries exist, as we believe it adds a small degree of rotational invariance to the network. In this respect, further experiments would be necessary to conclude whether rotaflip layers could also be beneficial in images without such symmetries.

The idea of performing regularization in FCNs --or in the convolutional part of CNNs-- has not been addressed as much as regularization in classic NNs \cite{Srivastava14}. Indeed, CNNs are already very efficient at reducing overfitting since they do not require as many parameters as in NNs. For instance, CNNs employed for classification usually contain the majority of parameters within the NN part and, thus, any effort to avoid overfitting is usually addressed in that section of the network. Alternatively, the increase in data samples, either by gathering more images or by artificial data augmentation, would suffice, particularly in FCN problems. However, this paradigm changed in recent years with the arrival of very deep CNNs, such as DenseNets \cite{Huang17} or ResNets \cite{He2016}, with dozens (sometimes hundreds) of convolutional layers and short-connections between them, such that gradient explosion but also overfitting were minimized. In both networks, they reduced the NN part to its minimum expression with simply two layers, heavily relying in the convolutional part instead of the NN section. Certainly, those networks have proven to be very robust, being considered the state-of-the-art in many CNN problems, yet there seems to be some overfitting to the training set due to the large number of parameters in the convolutional part.

In our first experiment (PatchCamelyon dataset with a DenseNet), we showed that the network was indeed overlearning the details of the training set, yet the inclusion of dropout layers decreased the accuracy. We believe that the bad performance of dropout was due to the complexity of the dataset and, thus, the need for the CNN to cover a very large number of complex features. In this respect, it seems clear that the addition of unrepresentative patterns by the dropout layers was indeed degrading the learning ability of the network. In contrast, dropout was a good regularizer in the segmentation of corneal endothelium images, probably because dropout mimics to certain extent the possible noisy patterns in the endothelial images and, thus, dropout contributes with helpful, representative examples. Furthermore, the endothelium problem was clearly a simpler one, where the challenge lied in covering the different types of artifacts that perturb the visual recognition of a grid of pseudo-hexagonal cells and not so much in describing the grid itself, which could be done with a limited number of features. This hypothesis was reinforced by the last experiment (the DenseUnet in the set of 1476 endothelial images) where the network did not overfit and the inclusion of dropout layers did not degrade the performance. Given the results of these experiments, we can conclude that dropout layers in between convolutional layers should be used with caution. 

In contrast, rotaflips have proven to be useful in all datasets but only with a small rate, quickly degrading the performance if the rate was increased beyond 10\%. This agreed with our hypothesis: knowing that any convolutional layer creates a more complex feature representation given several input features, a random rotational perturbation in only a very few input features would not distort the convergence of the convolutional filter and instead it could enforce the layer to encode the rotational transformations, whereas a random perturbation in many inputs would make the network unstable and unable to converge. Furthermore, we believe that the reduced number of perturbations (eight) is also suitable for this goal, as a much larger number of rotations (by doing interpolation) would create a more complex problem, although this hypothesis needs to be tested.

We also believe that the concept of rotaflip layers can be particularly beneficial in networks with short-connections, like DenseNets. Such networks do not rely in solving certain features in specific filters; instead, the concatenation of maps makes possible that a certain feature could be addressed by many filters along a dense block, probably each one addressing a small variant and thus making the network more robust; since we place the rotaflip layers in between a convolutional layer and a concatenation layer, we do not deprive the network of such tool and it becomes easier for the network to slowly learn the rotational transformations. Nevertheless, our experiments in U-net also proved that a network without such short-connections could also benefit from our method, but the rotaflip rate should be smaller.

Further experiments in similar datasets and networks could help to clarify the possibilities of our method, but these preliminary tests already suggest a clear potential.

\section{Conclusion}

The random rotation/reflection of a few feature maps (5--10\%) after each convolutional layer is a simple way to add regularization to very deep CNNs while providing a certain degree of rotational invariance. Our experiments showed that this novel method, named \textit{rotaflip}, can satisfactorily improve the performance in FCNs employed for semantic segmentation (U-nets) and in CNNs used for classification (DenseNets) in images with rotational symmetries, such as in medical images. Rotaflip layers can be interpreted as a substitution of dropout layers in the convolutional part of CNNs, achieving similar or better performance.


\bibliographystyle{splncs04}
\bibliography{eccv2020_rot}
\end{document}